\documentclass[a4paper,aps,prl,twocolumn,superscriptaddress]{revtex4}

\usepackage{graphicx} %[dvips]
\usepackage[usenames,dvipsnames]{xcolor}
\usepackage{tikz}
\usepackage{amsmath}
\usepackage{amssymb}
\usepackage{natbib}
\usepackage{color}
\usepackage{psfrag}
\usepackage{wasysym}
\usepackage{float}
\usepackage{epstopdf}
\usepackage[normalem]{ulem}
\usepackage{pgfplots}
\usetikzlibrary{pgfplots.groupplots}
\usepackage{textcomp}
\usepackage{hyperref}
\usepackage{soul}
\usepackage{amsmath}

\pgfplotsset{compat=newest}
\pgfplotsset{yticklabel style={text width=1.3em,align=right}, xticklabel style={text width=1.5em,align=center}}
\pgfplotsset{every axis legend/.append style={anchor=south east}}

\begin{document}
\title{Self-Ordering of Buckling, Bending, Bumping Beams}

\author{Arman Guerra}
\affiliation{Department of Mechanical Engineering, Boston University, Boston, Massachusetts 02215, USA}

\author{Anja Slim}
\affiliation{School of Mathematics, Monash University, Victoria 3800, Australia}
\affiliation{School of Earth, Atmosphere and Environment, Monash University, Victoria 3800, Australia}
\date{\today}

\author{Douglas P. Holmes}
\affiliation{Department of Mechanical Engineering, Boston University, Boston, Massachusetts 02215, USA}

\author{Ousmane Kodio} \email[]{kodio@mit.edu}
\affiliation{Department of Mathematics, Massachusetts Institute of Technology, Cambridge, Massachusetts 02139, USA}

\begin{abstract}

A collection of thin structures buckle, bend, and bump into each-other when confined. This contact can lead to the formation of patterns: hair will self-organize in curls; DNA strands will layer into cell nuclei; paper, when crumpled, will fold in on itself, forming a maze of interleaved sheets. This pattern formation changes how densely the structures can pack, as well as the mechanical properties of the system. How and when these patterns form, as well as the force required to pack these structures is not currently understood. Here we study the emergence of order in a canonical example of packing in slender-structures, \textit{i.e.} a system of parallel growing elastic beams. Using experiments, simulations, and simple theory from statistical mechanics, we predict the amount of growth (or, equivalently, the amount of compression) of the beams that will guarantee a global system order, which depends only on the initial geometry of the system. Furthermore, we find that the compressive stiffness and stored bending energy of this meta-material is directly proportional to the number of beams that are geometrically frustrated at any given point. We expect these results to elucidate the mechanisms leading to pattern formation in these kinds of systems, and to provide a new mechanical meta-material, with a tunable resistance to compressive force.

\end{abstract}


\maketitle

When thin structures pack, there is a competition between elasticity, which often encourages pattern formation and densification, and geometric constraints. 
For example, paper, when crumpled into a ball, forms complex three-dimensional swirls~\cite{cambou2011three,deboeuf2013comparative}, and DNA strands inserted into cell nuclei fold and pack into layers~\cite{smith2001bacteriophage, kindt2001dna}. In some cases, these densification processes are resisted by friction~\cite{alarcon2016self, alben2022packing} and geometrical incompatibilities in the deformation of the materials~\cite{andrejevic2021model,domokos2000constrained,roman1999buckling,pocheau2004uniqueness}. The formation of patterns has been studied thoroughly in cases where thin structures are adhered to a substrate~\cite{biot1937bending, dillard2018review, kodio2017lubricated, oshri2018delamination}, sheets are constrained in a ring~\cite{boue2006spiral, boue2007folding, adda2010statistical, alben2022packing}, and rods are inserted into a container~\cite{donato2003scaling, donato2007condensation, stoop2011packing, vetter2014morphogenesis, vetter2015packing}. However, the question of to what degree the rods and sheets will order themselves in these complex and random packing processes is still open.

\begin{figure}
\begin{center}
\vspace{1mm}
\includegraphics[width=0.98\columnwidth]{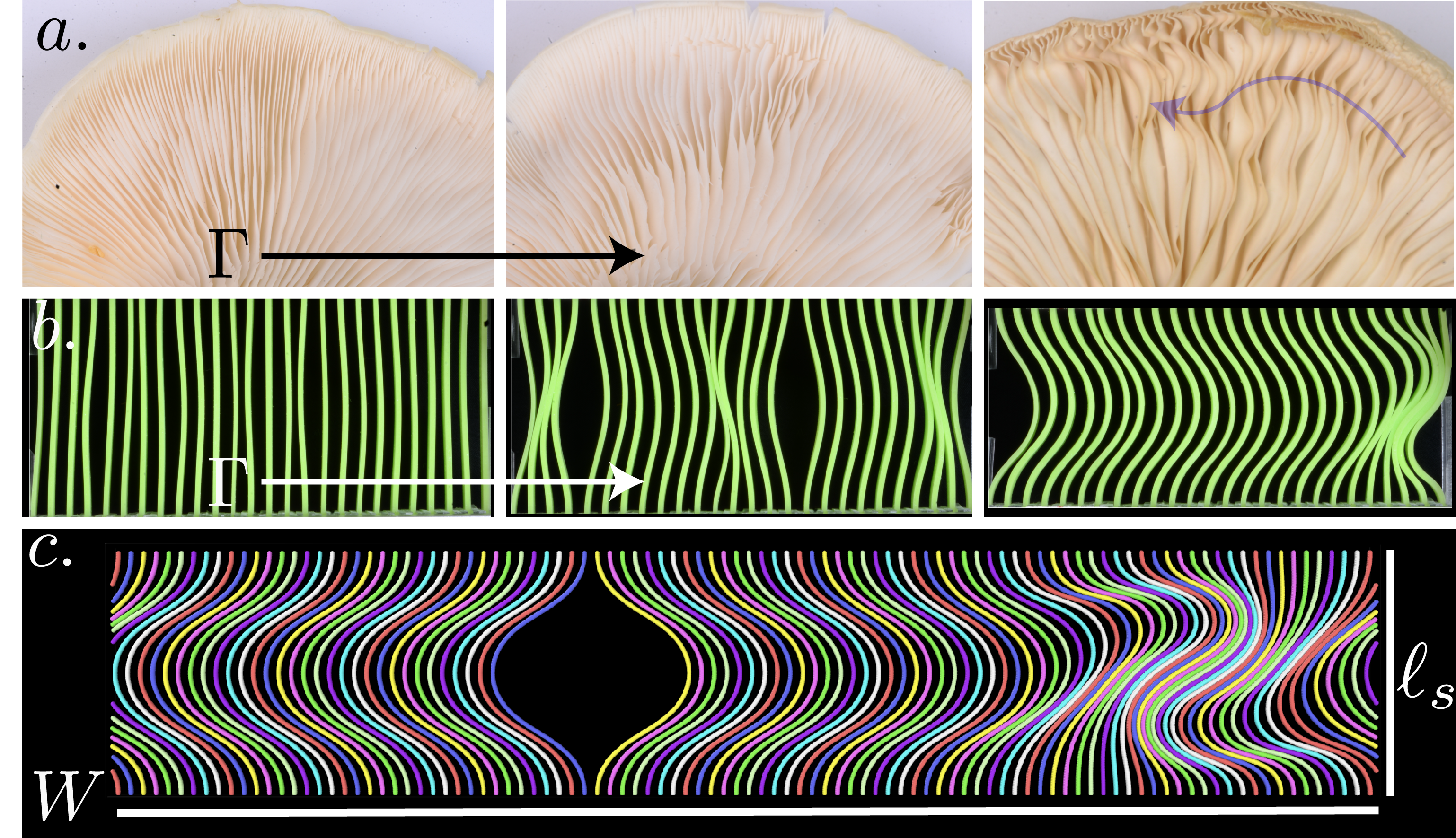}
\end{center}
\vspace{-5mm}
\caption{ (a) When an oyster mushroom dries, the gills become compressed along their length, causing them to buckle (middle) and then align (right). (b) Experimental observations of a similar phenomenon in a system of slender parallel plates compressed in an Instron. (c) A numerical experiment of the same system.
\label{fig1}}\vspace{-5mm}
\end{figure}

In structured arrangements of elastic beams, the competition between order and geometric frustration has been used to great effect in the design of materials with novel and programmed properties~\cite{florijn2014programmable, paulose2015selective, frenzel2016tailored, bertoldi2017flexible}. 
There are many models in statistical mechanics to rationalize the emergence of order, for example the Ising model~\cite{kang2014complex} for ferromagnetism and the Potts model~\cite{feynman2018statistical} for percolation, or even in the case of thin elastic structures \cite{bahri2022mechanical, hanakata2022anomalous, nelson2004statistical}. However, these models are insufficient to capture the ordering of beams because of the contact arising between them and the difficulty of finding the interacting forces between adjacent elements. For example, consider a simple 1-D version of the ordering of packed beams, the gills of a mushroom (see Figure~\ref{fig1}a). If the mushroom dries and shrinks, the gills will at first buckle, and then bump into each other. Reminiscent of the 1-D Ising model for magnetism, the ground state of this system occurs when all gills point in the same direction (shown in an analogous experiment in Figure~\ref{fig1}b, right). However, here there exists a hierarchy of disordered meta-stable states in the shallow-post-buckling regime, where the gills self-organize into ``clumps,'' and leave ``holes'' where they have separated (Figure~\ref{fig1}a, middle).

In this Letter, we consider a simplified version of this system: an array of $N$ parallel elastic beams confined to a vertical space $\ell_s$ that is shorter than their length $\ell$, and equally spaced inside of a box of width $W$, such that the distance between any two beams is $W/N$ (Figure~\ref{fig1}c). We characterize the control parameter of the system to be the confinement factor $\Gamma=\ell/\ell_s$. If the beams are initially perturbed in random directions, we observe behaviour reminiscent of a phase transition, where the initial disordered state (beams buckled in random directions) will gradually decay to the fully-ordered ground state (all beams aligned in one direction) as $\Gamma$ increases. We therefore ask the following two questions: First, when and how much will the beams align; can we predict how order will emerge? Second, how does the mechanical response of the system depend on the degree of order and its emergent topology; conversely could we use this emergent topology in the design of tunable-stiffness materials?

To investigate this alignment experimentally, we built a stiff acrylic two-piece mount that served to clamp a large number of slender elastic plates at two of their opposite edges. We inserted sheets of PVS (thickness $h=1.5$ mm, number $N=26$) into the mounts, and compressed the system uniformly with an Instron, which allowed us to measure $\Gamma$, as well as the force response of the arrangement of beams (Figure~\ref{fig1}b). We then perform non-thermal quenches on this ensemble of beams -- that is, we iso-statically compress the beams many times, each time manually biasing each beam by hand to initially buckle to the right or left based on a random coin-flip. Just as we see in the case of the mushroom (Figure~\ref{fig1}a), any adjacent beams buckling towards each other will eventually make contact and form clumps (Figure~\ref{fig1}b, middle). After enough confinement, these clumps become unstable and decompose, and the beams eventually all point in one direction (Figure~\ref{fig1}b, right). 

\begin{figure}
\begin{center}
\vspace{1mm}
\includegraphics[width=0.98\columnwidth]{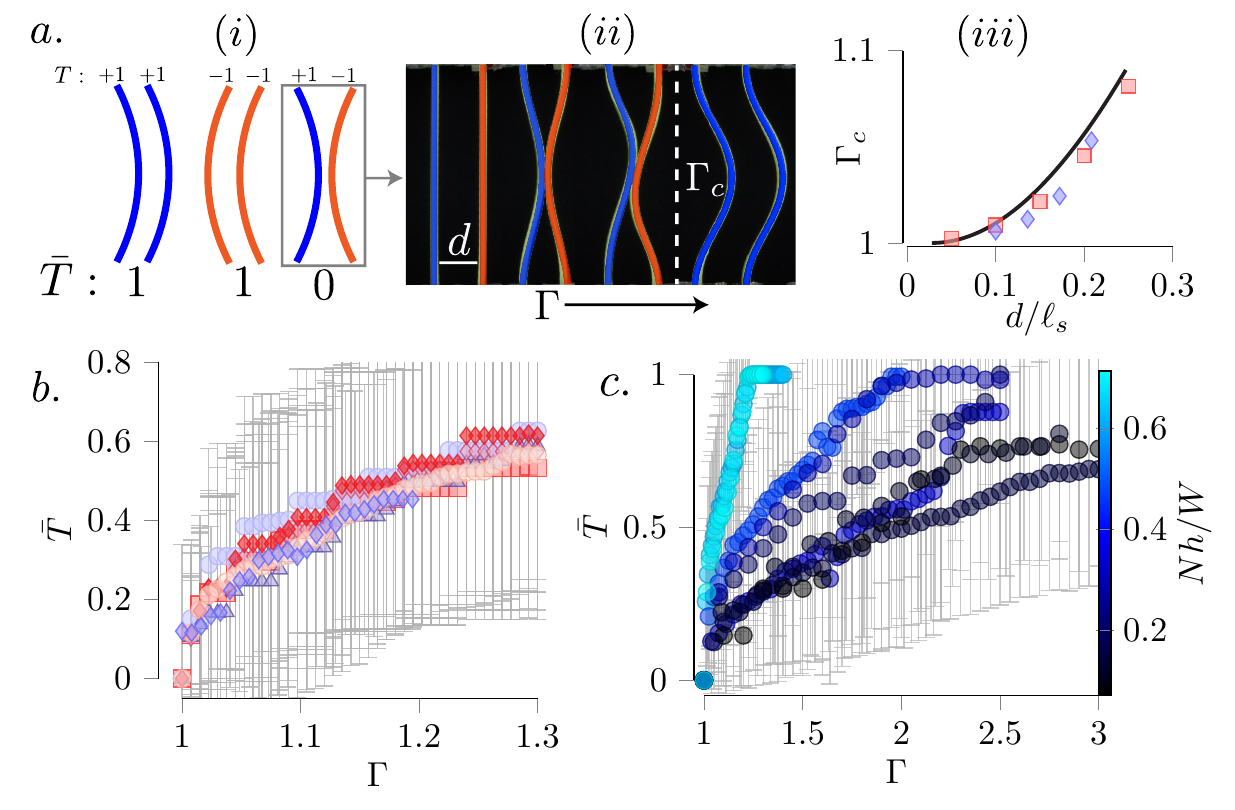}
\end{center}
\vspace{-5mm}
\caption{ The emergence of order for (a) two beams and (b $\&$ c) many beams. (a \textit{i.}) The tropism $T_i$ is $+1$ ($-1$) for a beam buckled to the right (left), which we indicate in blue (orange). The average tropism $\bar{T}$ (Equation~\ref{s}) quantifies the degree of system order. (a \textit{ii.}) Experiments (green) overlaid with simulations (blue and orange) of two beams which buckle towards each other. There are two meta-stable ``clumped'' $\bar{T} = 0$ states (vertically and rotationally symmetric) however when $\Gamma>\Gamma_c$, $\bar{T} = 1$. (a \textit{iii.}) $\Gamma_c$ increases with the normalized distance between two beams $d/\ell_s$ (experiments -- \protect\tikz \protect\draw[blue!70, fill=blue!30] (0.06cm,0cm) -- (0cm,0.1cm) -- (0.06cm,0.2cm) -- (0.12cm,0.1cm) -- (0.06cm,0cm);, simulations -- \protect\tikz \protect\draw[red!70, fill=red!30] (0,0) rectangle (0.2cm,0.2cm);, model -- black line (SI)) (b) $\bar{T}$  for experiments and simulations of many beams ($N=26$, $W=140$ mm, $\ell_s=54$ mm, $h=1.54$ mm, average $d=5.4$ mm, number of ensemble measurements: experiment -- 25, simulation -- 100) (c) $\bar{T}$ increases with $\Gamma$ with a rate highly dependent on $N$, $d$, $W$, $h$, and $\ell_s$.
\label{intromag}}\vspace{-5mm}
\end{figure}

We parametrize the direction of the buckling of each beam using what we call the ``tropism'' (analogous to the magnetism in the Ising model) where $T_i = 0$ when beam $i$ is un-buckled, and $T_i = +1$ ($-1$) if it is buckled to the right (left), as shown in Figure~\ref{intromag}a. To account for beams that are no longer in the first buckling mode because of contact with other adjacent beams or with the walls, we consider that a beam is buckled to the right (left) if the portion of the beam halfway up the box is farther to the right (left) than its ends. We can average the behavior of all beams into an overall system tropism

\begin{equation}
\bar{T} = \left| \frac{1}{N} \sum_{i=1}^{N} T_i \right|.
\label{s}
\end{equation}

Note that $ \bar{T}\approx 0$ when the beams are randomly directed, and $\bar{T} \approx 1$ when the beams are all aligned (Figure~\ref{fig1}b, right). We plot $\bar{T}$ for increasing $\Gamma$ for a fixed beam spacing, number, thickness, and initial length in Figure~\ref{intromag}b (\protect\tikz \protect\draw[blue!60, fill=blue!40] (0.06cm,0cm) -- (0cm,0.1cm) -- (0.06cm,0.2cm) -- (0.12cm,0.1cm) -- (0.06cm,0cm);), and find indeed a gradual increase in the average order of the system. We can compare this with the behavior of two beams buckling towards each other (Figure~\ref{intromag}a \textit{ii} and \textit{iii}) and find that for the many-beam case, the clumps remain stable (and therefore $\bar{T}<1$) at much higher $\Gamma$ for the same normalized beam spacing $d/\ell_s$ (Figure~\ref{fig1}c).

To investigate how the observed behavior depends on the friction between the beams, the boundary conditions, the variation in distances between adjacent beams, and any imperfections in the experimental set-up (including electrostatic interactions between neighboring plates and gravity), we supplemented our physical experiments with numerical ones using the Large-scale Atomic/Molecular Massively Parallel Simulator (LAMMPS~\cite{thompson2022lammps}, with visualizations in Ovito~\cite{stukowski2009visualization}) using a similar protocol as~\cite{guerra2021emergence}, which is meant to replicate the physics of elastic beams. This consisted of simulating a large number of point particles with diameter $h$ equal to the thickness of the beam with nearest-neighbor potentials $U=Y(r-r_0)^2+B(1+\cos{\theta})$ where $Y=E_l \pi h/8$ is the stretching modulus, $B=E_l \pi h^3/64$ is the bending modulus, $E_l$ is the Young's modulus of the beams, $r$ and $r_0$ are the current and equilibrium nearest-neighbor distances respectively, and $\theta$ is the angle between any three adjacent beam particles. We note that $r_0$ is not necessarily equal to $h$, in fact in all simulations we use $r_0 = h/4$ to allow the beams to slide smoothly against one another. 

In Figure~\ref{intromag}b, we replicate the geometry of our physical experiments and plot the ensemble behavior of $\bar{T}$ for these simulations (\protect\tikz \protect\draw[Blue!70, fill=Blue!30] (0,0) -- (0cm,0cm) -- (0.1cm,0.2cm) -- (0.2cm,0cm) -- (0cm,0cm);). 
We include simulations with varying inter-beam friction, where simulations with coefficient of sliding friction set to $\mu_s=0$ (frictionless, \protect\tikz \protect\draw[red!80, fill=red!40] (0,0) rectangle (0.2cm,0.2cm);), $\mu_s=1$ (rubber-like, \protect\tikz \protect\draw[purple!70, fill=purple!30] (0.06cm,0cm) -- (0cm,0.1cm) -- (0.06cm,0.2cm) -- (0.12cm,0.1cm) -- (0.06cm,0cm);), $\mu_s=2$ (highly frictional, \protect\tikz \protect\draw[red!100, fill=red!100] (0.06cm,0cm) -- (0cm,0.1cm) -- (0.06cm,0.2cm) -- (0.12cm,0.1cm) -- (0.06cm,0cm);). 
The effect of boundary conditions is investigated by performing simulations with $\mu_s=0$ in which we vary the beam end boundary conditions (pin-pin, \protect\tikz \protect\draw[blue!30, fill=blue!20] (0,0) circle (.55ex);), and the vertical box edge boundary conditions (periodic, \protect\tikz \protect\draw[red!70, fill=red!30] (0,0) rectangle (0.2cm,0.2cm);). 
Additionally, we performed simulations where the distances between the beams was uniformly randomly distributed with a lower bound of $1.2 h$ and an upper bound such that the average distance between the beams is the same as the model experiment (\protect\tikz \protect\draw[Red!30, fill=Red!10] (0,0) circle (.55ex);). We find good agreement between experiments and simulations, and furthermore the tropism $\bar{T}$ is statistically independent of the degree of friction, boundary conditions, and beam spacing variability. Since the tropism does not depend on any of these choices, we can select one simulation protocol and vary the geometric parameters, with the expectation that the results apply broadly. Hence, without loss of generality, we perform simulations with clamped-clamped, evenly spaced beams without friction in a periodic box. We vary the beam density as well as the number of beams, and plot $\bar{T}$ as a function of $\Gamma$ in Figure~\ref{intromag}c, and find that $\bar{T}$ strongly depends on the geometric parameters of the box. 

%, and the final direction of all the beams is simply the direction that the majority of the beams buckled initially.

% \textcolor{red}{check, also say not necessarily the case for randomly spaced beams}.

\begin{figure}
\begin{center}
\vspace{1mm}
\includegraphics[width=0.98\columnwidth]{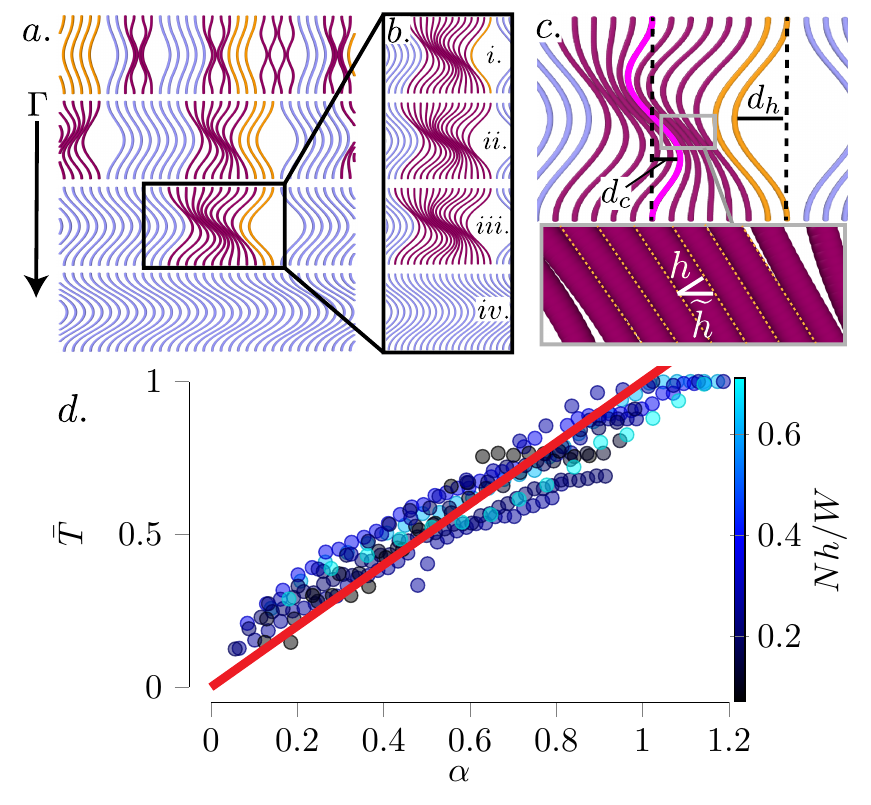}
\end{center}
\vspace{-5mm}
\caption{ Using the mechanism behind alignment, we can predict the tropism. (a) Simulations at increasing $\Gamma$ from top to bottom. Beams which are part of a clump are colored purple, and all others are colored by their tropism ($T_i=1 \rightarrow$ blue, $T_i=-1 \rightarrow$ orange). (b) A step-by-step illustration of clump-hole annihilation: (\textit{i}) before the clump and hole meet, the lateral force on the clump is balanced by beams on both sides. (\textit{ii}) When the final beam between the hole and the clump meets the clump, the force on the clump is still balanced. (\textit{iii}) The clump becomes unstable when it is touched by a beam on the left which does not have a counterpart on the hole side. (\textit{iv}) The clump quickly moves into the hole, and all the beams become aligned. (c) Illustration of the parameters in our mathematical model. (d) If we compare the sum of the distances defined in (c) with the maximum possible distance between a clump and a hole, the ensemble value of $\bar{T}$ falls onto a single curve (red). Error bars are the same as in Figure~\ref{intromag}c, left off for clarity.
\label{collapse}}\vspace{-5mm}
\end{figure}

In Figure~\ref{collapse}a we show the evolution of a specific set of LAMMPS simulated beams as $\Gamma$ increases. Beams that are in contact with a neighbor are colored red and non-contacting beams are colored by their tropism ($1\rightarrow$ blue, $-1\rightarrow$ orange). Adjacent beams that buckle toward each other form clumps, while beams that buckle away form holes. We note that these clumps also exist in the case where the beams are randomly distributed in the box; they are not just a feature of the evenly spaced beams. We see that as $\Gamma$ increases, the number of clumps and holes decrease, leading to increasingly aligned beams. We can uncover the mechanism by which the beams align by zooming in on the instance shown in Figure~\ref{collapse}b. In Figure~\ref{collapse}b\textit{i} we see that the clump and hole are separated by one beam that is not in contact with the clump, and that the force on the clump is balanced, that is, there are as many beams on the right buckled towards the left as there are beams on the left buckled towards the right. In Figure~\ref{collapse}b\textit{ii}, the beam on the right of the clump has made contact with the clump, but so has a beam on the left of the clump, so again the clump is balanced. In Figure~\ref{collapse}b\textit{iii} however, an additional beam makes contact with the left of the clump, but is not balanced by the beam that would have otherwise been on the right of the clump to balance it. The force on the clump is therefore unbalanced, and the clump rushes into the hole, leading to clump-hole annihilation, and the alignment of all the beams (Figure~\ref{collapse}b\textit{iv}).

We can use this mechanism to predict $\bar{T}$ as a function of $\Gamma$. We know that $\bar{T} = 1$ when the final clump ``meets'' and annihilates with the final hole. For simplicity we will consider the case where the edges of the box are periodic, and as such the maximum possible distance between the final clump-hole pair is $W/2$. What does it mean for a clump and a hole to meet? A clump and a hole meet when all of the beams between them are in contact. In this case, the space between the center of the hole and the center of the clump comprises three parts as illustrated in Figure~\ref{collapse}c. The first is the horizontal extent of the hole, which we call $d_h$. If we approximate the shape of a beam as a triangle, using the Pythagorean theorem, $d_h \approx \frac{1}{2}\ell_s\sqrt{\Gamma^2-1}$. The second is the horizontal deflection of the center beam of the clump, which we call $d_c$. This has the approximate shape of a mode-2 buckled elastica, so we can approximate this as $d_c \approx d_h/2$. The third is the sum of the thicknesses of the beams between the clump and the hole. The number of such beams $N_b$ stays the same as the beams grow, but the horizontal thickness $\widetilde{h}$ of the beams change, as shown in Figure~\ref{collapse}c. If we keep our triangular approximation for the shape of the beams, we get that $\widetilde{h} \approx h\Gamma$ and therefore the additional space that each beam takes up is $\widetilde{h}-h=h(\Gamma-1)$.

Since the largest possible distance between a clump and a hole is $W/2$, therefore the maximum amount of empty space between the clump and hole is the distance between the clump and the hole minus the thickness of the beams between them, or $W/2-Nh/2$. We would expect that when the beams grow and this space is taken up, that is, $d_h+d_c+N_bh(\Gamma-1) = W/2-Nh/2$, there will be nothing separating the farthest possible clump and hole, they will make contact and annihilate, and $\bar{T} \rightarrow 1$. In the case where the last clump and hole are as far apart as they can be, $N_b = N/2$, so the fraction of the horizontal space between the clump and hole taken up by the beams, which we will call the ``porosity'' $\alpha$, is $\alpha = [d_h+d_c+N(\widetilde{h}-h)/2]/(W/2-Nh/2)$. In Figure~\ref{collapse}d, we plot $\bar{T}$ against this porosity and find that, as we expect, for all geometries, $\bar{T} \approx 1$ when $\alpha = 1$. More than that however, it seems that the tropism is {\em approximately equal} to $\alpha$, that is,

\begin{equation}
	\bar{T} \approx \alpha = \frac{\frac{1}{2} (\Gamma -1) h N+\frac{3}{4} \ell_s \sqrt{\Gamma^2-1} }{W/2-Nh/2}.
	\label{hello}
\end{equation}

\noindent This comes from the fact that as the beams grow, the number of clumps and holes which make contact and annihilate is proportional to the space that the beams take up. We note that, as can be seen from Figure~\ref{collapse}d, even when $\alpha = 1$, there are some cases where $\bar{T}$ is slightly less than 1. We expect that this comes from errors in our approximations of $d_h$, $d_c$, and $\widetilde{h}$.

Earlier we showed that $\bar{T}$ does not significantly depend on the boundary conditions of the system, the friction between beams, or the variation in the distance between beams. We can now see why: the boundary conditions of the beams do not significantly affect the validity of our approximations for the maximum beam deflection, and the box walls and variability in inter-beam distance affect neither the average distance nor (in the case of an ensemble average of experiments) the average number of beams between the holes and clumps. In our simulations we also did not see an effect of friction on the shape of the beams in the clumps, indicating that the beams in this specific geometry do not need to slide past each other for order to emerge.

 \begin{figure}
\begin{center}
\vspace{1mm}
\includegraphics[width=1\columnwidth]{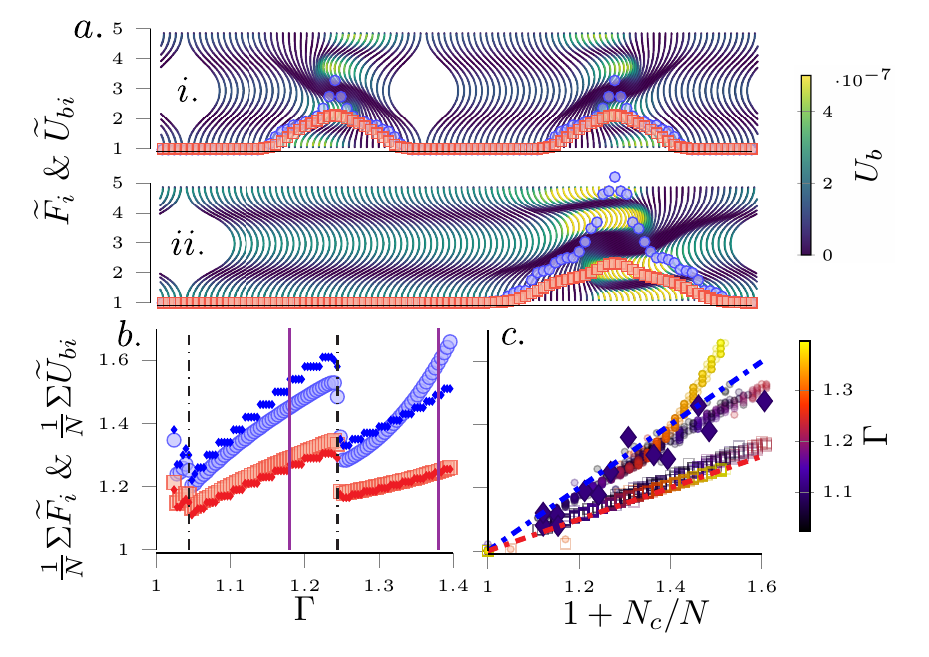}
\end{center}
\vspace{-5mm}
\caption{ The compressive stiffness of the system is a function of the degree of order. (a) Snapshots of a simulation at $\Gamma=1.18$ (\textit{i}) and $\Gamma=1.38$ (\textit{ii}), superimposed with the vertical compressive force ($F_i$, blue) and bending energy ($U_{bi}$, red) of each beam normalized by the force and bending energy of a beam which is not part of a clump ($F_1$, $U_{b1}$), showing that the clumps have a higher resistance to compression, and a larger average curvature than the non-clumped regions. (b) The total compressive force and bending energy ($F$, $U_b$) of the same system of beams as a function of $\Gamma$, normalized by the compressive force and bending energy of the system if all of the beams were buckled in the same direction. The circles and squares are experiments and simulations with the same initial conditions, and the diamonds are the predictions from our approximated model (Equation~\ref{foob}). Vertical lines show places where a clump and hole annihilate. (c) Normalized force and bending energy against $1+N_t/N$ together with our predictions from Equation~\ref{foob} (simulation $U_b$ -- squares, simulation $F$ -- circles, experiment $F$ -- diamonds).
\label{fub}}\vspace{-5mm}
\end{figure}

Now that we understand when the beams will become ordered, we turn to our second question: how do the beams respond to the imposed compression when they are not ordered? This is analogous to why it requires less force to confine paper to a target volume through folding than through crumpling~\cite{deboeuf2013comparative}: a larger degree of geometrical frustration in thin structures often leads to more stored energy~\cite{sadoc1999geometrical}. We can observe this directly in our system. In Figure~\ref{fub} we plot the normalized compressive force $\widetilde{F}_i = F_i/F_1$ of each beam in a simulation, where $F_i$ is the compressive force applied to beam $i$, and $F_1$ is the compressive force applied to a single beam with the same boundary conditions, compressed to the same $\Gamma$ and no neighboring beams. We simultaneously define and plot the normalized bending energy $\widetilde{U}_{bi} = U_{bi}/U_{b1}$ in the same way, where $U_{bi}$ is the bending energy of beam $i$, and $U_{b1}$ is the corresponding bending energy of the beam from which we derive $F_1$. We superimpose the normalized compressive force graph onto the simulated configurations and colored beam particles by their bending energy. For beams that are not in clumps, and therefore have no contacts, $\widetilde{F}_i = \widetilde{U}_{bi} = 1$. In contrast, beams that are in contact with other beams in a clump and are geometrically frustrated~\cite{sadoc1999geometrical} have a higher $\widetilde{F}_i$ and $\widetilde{U}_{bi}$, as they cannot find their lowest-energy state. These results indicate that our system is behaving like a mechanical meta-material, an object that gets its properties not only from the material that it is made of, but also its internal geometry~\cite{bertoldi2017flexible}.

Previously, we noted that the beam at the center of a clump has the approximate shape of a mode-2 elastica. We would expect that $U_{b} \propto n$ where $n$ is the mode number of the beam (since doubling the mode number on average doubles the curvature in the beam), so for the central beam in the clump we might expect that $\widetilde{U}_{bi} \approx 2$. This approximation is confirmed in Figure~\ref{fub}a. Even for very large $\Gamma$ we find that the central beam in any clump has a normalized bending energy of approximately 2. To estimate the normalized compressive force of a beam ($F_i$) at the center of a clump, we perform additional simulations of a single beam, where the slope at the vertical center of the beam was forced to match the slope derived from our triangular approximation of the beam shape. In these simulations we find that $\widetilde{F}_{i} \approx 3$ for the approximated shape. In Figure~\ref{fub}a, we find that the maximum value of $\widetilde{F}_{i}$ is indeed approximately $3$ for most of the simulation, but increases for very large $\Gamma$, which we expect comes from a deviation of the true beam shape from our approximation.

From the graphs in Figure~\ref{fub}a we see that the normalized bending energy and compressive force of the beams in the clump seems to increase linearly from $\approx 1$ for the beams at the edge of the clump, to the maximum value at the clump center. If we use our earlier approximations ($\widetilde{U}_{bi} \approx 2$ and $\widetilde{F}_{i} \approx 3$), we might expect that, on average, a beam which is in a clump has $\widetilde{U}_{bi} \approx 1.5$ and $\widetilde{F}_{i} \approx 2$. Hence, we would expect that the total force and bending energy in our arrangement of beams is

\begin{equation}
\begin{split}
& F = F_1(N+N_c),   \\ 
& U_b = U_{b1}(N+N_c/2),
\label{foob}
\end{split}
\end{equation}

\noindent where $N_c$ is the number of beams that are in a clump. In Figure~\ref{fub}b, we plot the total normalized force $F/(NF_1)$ (\protect\tikz \protect\draw[blue!70, fill=blue!30] (0,0) circle (.55ex);)
 and bending energy $U_b/(NU_{b1})$ (\protect\tikz \protect\draw[red!70, fill=red!30] (0,0) rectangle (0.2cm,0.2cm);) of the specific simulation run pictured in Figure~\ref{fub}a, along with their estimated values from the number of clumped beams as given by Equation~\ref{foob} and find good agreement, except at high $\Gamma$, when the normalized force estimate starts to fail as we expected. In Figure~\ref{fub}c, we plot $F/(NF_1)$ (circles) and $U_b/(NU_{b1})$ (squares) against $1+N_c/N$ and find that for a wide variety of box geometries, $U_b/(NU_{b1})$ collapses to $ (1+N_c/N)/2$ for all data, and $F/(NF_1)$ collapses to $1+N_c/N$ for all data except for that at very high $\Gamma$.

In this work we have studied a system of many parallel, growing beams clamped at their ends and confined in a box. We have found that, as they grow, clumps of beams that form when the beams first buckle eventually meet with holes where the beams have separated, and the beams begin to order themselves. The degree of order (which we parameterize as the tropism $\bar{T}$) is proportional to the fraction of the initially empty horizontal space that the beams take up as they grow. Our results indicate that, for parallel growing beams constrained in a rectangular box, at high enough $\Gamma$, $\bar{T}\rightarrow 1$, that is, we had no results indicating that the beams would ``jam,'' or lock in place. We expect that with a different geometry, however, these geometrically frustrated states could become locked in place by the inter-beam friction, similar to interleaved books~\cite{alarcon2016self}, bird nests~\cite{weiner2020mechanics}, and a beam injected into a tube~\cite{van1993validation, gao2009effects, miller2015extending}. We also probed the mechanical properties, and found that the compressive stiffness of the metamaterial made up of these buckling beams is proportional to the number of beams in a clump. This provides a system wherein the stiffness can be manually or automatically tuned by forcing beams into our out of a clump while the metamaterial is under load. We might expect that, similar to the tropism, friction could play a major role in the mechanical properties of this kind of system, as it does in many cases of slender objects in contact~\cite{poincloux2021bending}, creating an additional route for potentially novel functionalities.

We have, so far, not considered beams which have any intrinsic thermal motion~\cite{nelson2004statistical}, adhesion~\cite{elder2020adhesion, cranford2022crumpled}, or long-range potentials, all of which would affect the ordering, and mechanical properties of the system. We also note that, as the beams are compressed, a transverse force (the compression) turns into a longitudinal transfer, namely the redirection of the beams. This could provide a method to redirect and control mechanical waves~\cite{wu2020mechanical, packo2021metaclusters}.

\section{Acknowledgements}

We thank Abigail Plummer, Harold Park, and Dominic Vella for helpful discussion. We also gratefully acknowledge the financial support from DARPA (\#HR00111810004) and from NSF CMMI--CAREER through Mechanics of Materials and Structures (\#1454153), and the computing resources of the Boston University Shared Computing Cluster.

\section{SI}

\subsection{Critical Confinement Factor of Two Contacting Beams}

In Figure~\ref{intromag} we show simulations and experiments of two beams which come into contact and then, at a critical confinement factor ($\Gamma_c$), one beam wins, and the other snaps-through such that both beams are pointing in the same direction. We might expect that when the space between the two beams is taken up by the extension of a beam ($d_h$) and the thickness of the beams $\widetilde{h}$, the beams will align. In other words, the beams will align when the spacing between the beams is $d_h + \widetilde{h} = \frac{1}{2} \ell_s \sqrt{\Gamma_c^2-1}+h\Gamma_c$. We plot this in Figure~\ref{intromag}a \textit{iii}. (black line) and find good agreement with simulated (\protect\tikz \protect\draw[red!70, fill=red!30] (0,0) rectangle (0.2cm,0.2cm);) and experimental (\protect\tikz \protect\draw[blue!70, fill=blue!30] (0.06cm,0cm) -- (0cm,0.1cm) -- (0.06cm,0.2cm) -- (0.12cm,0.1cm) -- (0.06cm,0cm);) values.

%%%% Bibliography

\end{document}